# Harmonic-Balance Based Power Flow and ZVS Analysis of a Quad-Active Bridge DC-DC Converter

Ezekiel Olayiwola Arogunjo
Department of Electrical and Computer Engineering,
Tennessee Tech University
Cookeville, USA
eoarogunjo42@tntech.edu

Nnadi Olivia
Department of Electrical and Computer Engineering,
Tennessee Tech University
Cookeville, USA
onnadi42@tntech.edu

Joseph Olorunfemi Ojo
Department of Electrical and Computer Engineering,
Tennessee Tech University
Cookeville, USA
jojo@tntech.edu

*Abstract*—The power flow control of multi-active bridge converters requires a comprehensive steady-state analysis of the converter and the determination of conditions for zero voltage switching of all switches in the converter that result in minimum switching loss. This paper aims to model and carry out the power flow and Zero Voltage Switching (ZVS) analysis of Quad-active-bridge (QAB) dc-dc converter. The dynamic as well as the steady state analysis of the converter were carried out, thereby determining the phase shifts required to meet commanded load powers. The full equivalent circuit model of the converter which include winding resistances and magnetizing inductance is used rather than the popular lossless star-equivalent circuit model that may introduce significant model error in the converter's analysis. The conditions which ensure the converter working in ZVS mode are determined and experimentally verified.

## I. INTRODUCTION

High penetration of renewable energy and electric vehicle systems have increased interests in the design of advanced power electronic converters. Recently, the requirements for strong integration of loads and various power sources and storage systems in renewable energy and electric vehicle charging systems have necessitated the designs of modular architecture in power converter systems.

For the proposed QAB, as described in [1], Each of the bridges are connected to the primary side of two-winding independent transformers through an auxiliary energy transfer inductors

Several authors have carried out the modeling and analysis of a multi-active dc-dc converter. In carrying out the power flow and ZVS analysis of a triple active bridge, many authors convert it from its delta topology to a star structure so as to apply the methods of analysis of a DAB to the TAB [2], [3]. For instance, Purgat et al, carried out the power flow and ZVS analysis of TAB by converting it to a star structure. This may however complicate the analysis of a multi active bridge with more than 3 ports [4]. Bandyopadhay et al. presented a decoupling control strategy based on a linear active disturbance rejection control. They carried out the power flow analysis using the star equivalent model and determined the ac link voltage of the converter through the superposition theorem [5]. Ortega et al, carried out the steady-state analysis of a multi-active bridge dc-dc converter and deployed an iterative algorithm to calculate the phase shifts to determine the demanded power at the ports [1]. Naseem et al, presented an asymmetrically configured quad active bridge dc-dc converters based on current balancing coupled inductor. In the modeling and analysis of the converter, they also neglected the resistance and magnetizing inductance of the converter [6]. Falcones also modeled and analyzed the quad active bridge using the star equivalent circuit, neglecting the magnetizing inductance and winding resistances of the high frequency transformer in the process [7]. All the authors reviewed assumed a lossless equivalent circuit model where the algebraic sum of the power at the ports is zero. This may introduce significant model errors in high power applications.

This article considers the power flow analysis of a Quad-active Bridge dc-dc converter using its full equivalent circuit (including the leakage and magnetizing inductances of its high frequency transformers and winding resistances). Harmonic balance technique is deployed to carry out the steady-state analysis of the converter. This method allows for an easier analysis of a more complex converter compared to the actual state-space model of the converter. The ZVS analysis of the converter is also carried out considering the total output power and load voltages.

## II. CONVERTER DESCRIPTION AND MODELING

The proposed four-port multi-active bridge or Quad-active bridge converter (QAB) as shown in Figure 1 consists of four full bridges, each connected to the primary winding of a high frequency transformer. The four high frequency transformers are connected in parallel through the secondary windings. Ports 1 and 3 of the QAB are the source ports while, Ports 2 and 4 are the load ports.

The full bridges that make up the ports are operated by the single-phase shift (SPS) control. As shown in Figure 3, the output voltage of each bridge is a square-wave (50% duty cycle), shifted by $\phi_i = \delta_i\pi$, $i = 1,2,...,4$, where $\delta_i \in [-1,1]$ are the converters' phase shift ratios. The angle shifts enable control of the power flow among the component full bridges in the Quad Active Bridge dc-dc converter.

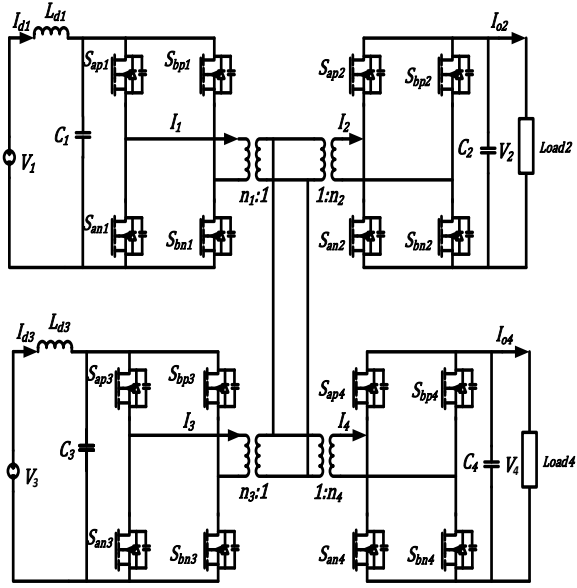

Fig. 1. Quad-active bridge dc-dc converter topology

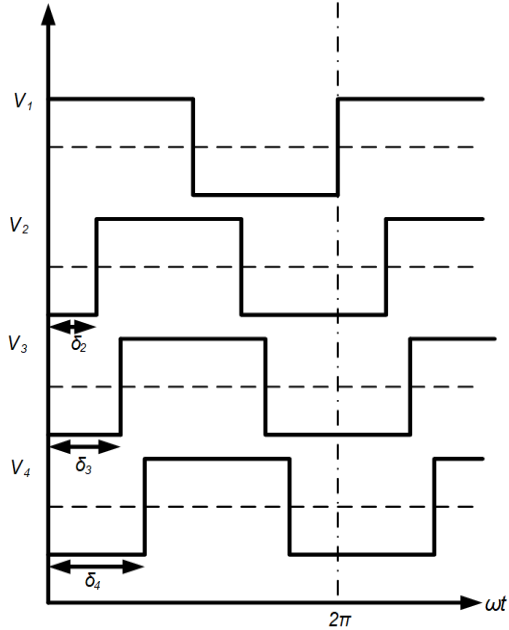

Fig. 2. Quad-active bridge dc-dc converter phase shift description

The converter's dynamic equations for the port currents are derived from the mesh and nodal analysis of the converter's equivalent circuit shown in Figure 3. All secondary transformer parameters have been referred to the primary sides. The model is given in state-space form as:

$$\dot{x} = Ax + Bu \quad (1)$$

Where, $x = [I_{11} \ I_{21} \ I_{22} \ I_{31} \ I_{32} \ I_{42} \ I_s]^T$,

$u = [v_1 \ v_2 \ v_3 \ v_4 \ 0 \ 0 \ 0]^T$,

$A = -A_L^{-1} A_R, \quad B = A_L^{-1}$

$$A_L = \begin{bmatrix} A_{L11} & -L_{m1} & 0 & 0 & 0 & 0 & -L_{m1} \\ 0 & L_{m2} & A_{L23} & 0 & 0 & 0 & 0 \\ 0 & 0 & 0 & A_{L34} & -L_{m3} & 0 & 0 \\ 0 & 0 & 0 & 0 & L_{m4} & A_{L46} & L_{m4} \\ L_{m1} & A_{L52} & L_{m2} & 0 & 0 & 0 & -(L_{m1}+L_{12}) \\ 0 & 0 & 0 & L_{3m} & A_{L65} & L_{4m} & -(L_{m4}+L_{41}) \\ 0 & A_{L72} & 0 & 0 & A_{L75} & -L_{m4} & L_{41}+L_{m4} \end{bmatrix}$$

$A_{L11} = L_{11} + L_{m1}, \ A_{L23} = -(L_{m2} + L_{22})$

$A_{L34} = L_{31} + L_{m3}, \ A_{L46} = L_{41} + L_{m4}$

$A_{L52} = -(L_{12} + L_{m1} + L_{m2} + L_{21}), \ A_{L72} - (L_{21} + L_{m2})$

$A_{L65} = -(L_{m3} + L_{m4} + L_{32} + L_{41}), \ A_{L75} = -(L_{m4} + L_{42})$

$$A_R = \begin{bmatrix} R_{11} & 0 & 0 & 0 & 0 & 0 & 0 \\ 0 & 0 & R_{22} & 0 & 0 & 0 & 0 \\ 0 & 0 & 0 & R_{31} & 0 & 0 & 0 \\ 0 & 0 & 0 & 0 & 0 & -R_{42} & 0 \\ 0 & -(R_{12}+R_{21}) & 0 & 0 & 0 & 0 & -R_{12} \\ 0 & 0 & 0 & 0 & -(R_{32}+R_{41}) & 0 & -R_{41} \\ 0 & R_{21} & 0 & 0 & R_{41} & 0 & R_{41} \end{bmatrix}$$

The currents in (1) are as defined and indicated on the equivalent circuit diagram. The ac side current of the full bridges are given by:

$$I_1 = I_{11}, \ I_2 = I_{22}, \ I_3 = I_{31}, \ I_4 = I_{42}$$

$v_1$ and $v_3$ are the output ac voltages and $v_2$ and $v_4$ are the input ac voltages of the quad active bridge dc-dc converters. $A_L$ and $A_R$ are the network's inductance and resistance matrices respectively. Using (1) and the parameters in Table 1, the converter is simulated in the MATLAB/SIMULINK software. The simulation results based on the state space (SS) (in blue) and harmonic balance (HB) (fundamental component is in red) model are compared in Figures 4, 5 and 6. The output ac voltage and current waveform of H-bridges are shown in Figures 4 and 5. The simulation result of the ac link voltage, $v_{ac}$ and current $I_s$ as indicated in Figure 3 are shown in Figure 6. From this result, it is observed that the result of the full bridges output ac currents $I_1, I_2, I_3, I_4$ and the ac link current $I_s$ for the state-space modeling method and the fundamental component of the HB method are very close. Hence, the fundamental component of the model from the HB technique is a good approximation of the state-space model.

### III. POWER FLOW ANALYSIS

Since harmonic balance technique (HB) is a more tractable method of analysis of multi-active bridge dc-dc converters, it is used to derive the current and power equations at the QAB ports. The first harmonic approximation of the bridge output voltages and current are given by:

$$v_i = \frac{4V_i}{\pi} \cos\left(\omega_s t - \phi_{v_i} - \frac{\pi}{2}\right), \quad (2)$$

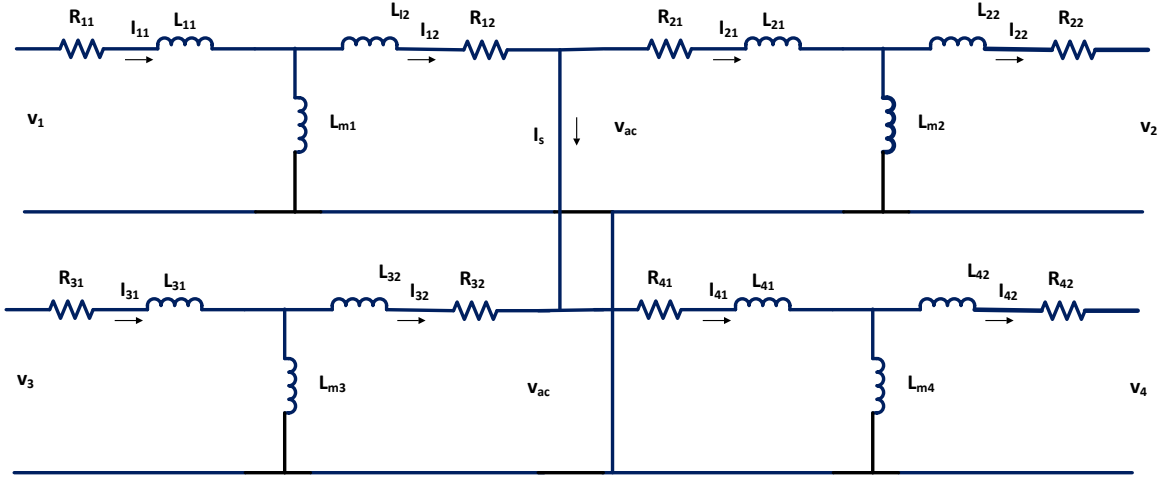

Fig. 3. Quad-active bridge dc-dc converter equivalent circuit

Where $V_i$ s are the input dc voltages, $\omega_s$ is the QAB switching frequency in radian per second, $\phi_{v_i}$ are the phase shift angles of the bridges output voltages.

Hence, the impedance matrix of the converter derived from (1) is given by:

$$Z = j\omega_s A_L + A_R \quad \to Y_f = Z^{-1} \quad (3)$$

Where $Y_f$ is the admittance matrix of the network in Figure 3. The rows of the $4 \times 4$ admittance matrix $Y$ relating the ports currents and voltages is extracted from $Y_f$ as:

$$y_1 = Y_f(1), \quad y_2 = Y_f(3), \quad y_3 = Y_f(4), \quad y_4 = Y_f(6)$$
$$Y = [y_1 \; y_2 \; y_3 \; y_4]^T$$

Hence the fundamental components of the port currents are given as:

$$I_i = \sum_{j=1}^{n} Y_{ij} v_j \quad (4)$$

The popular method of solving for the phase shift angles to meet demanded load power is to assume a lossless star-equivalent circuit of the QAB and specify all ports' power to algebraically add up to zero. In high power applications, this assumption may introduce a significant error in the analysis. For instance, with a winding resistance as low as 0.005 ohms and using the transformer parameters in [1], the copper loss is as high as 9.6kW.

In this article, the winding resistance of the transformers are included in the model. As a result, the algebraic sum of the ports' powers will not be zero. To determine the phase angles for the converter, only the powers at the load ports are specified. The power flow between the source ports is equated to zero to prevent power circulation between sources. converter 1 is assumed as the reference converter. The phase shifts of the output ac voltages of converters 2, 3 and 4 from the output ac voltage of converter 1, are solved by numerical methods. The equation for the converters 1 and 3 (source converters) input powers and Converters 2 and 4 (load converters) output powers are given as:

$$P_1 = \frac{8V_1^2 G_{11}}{\pi^2} + \frac{8V_1 V_2 G_{12}}{\pi^2} \cos(\phi_{v_2} - \phi_{v_1}) + \frac{8V_1 V_2 B_{12}}{\pi^2} \sin(\phi_{v_2} - \phi_{v_1}) + \frac{8V_1 V_3 G_{13}}{\pi^2} \cos(\phi_{v_3} - \phi_{v_1}) + \frac{8V_1 V_3 B_{13}}{\pi^2} \sin(\phi_{v_3} - \phi_{v_1}) + \frac{8V_1 V_4 G_{14}}{\pi^2} \cos(\phi_{v_4} - \phi_{v_1}) + \frac{8V_1 V_4 B_{14}}{\pi^2} \sin(\phi_{v_4} - \phi_{v_1}) \quad (5)$$

$$P_3 = \frac{8V_3^2 G_{33}}{\pi^2} + \frac{8V_3 V_1 G_{31}}{\pi^2} \cos(\phi_{v_1} - \phi_{v_3}) + \frac{8V_3 V_1 B_{31}}{\pi^2} \sin(\phi_{v_1} - \phi_{v_3}) + \frac{8V_3 V_2 G_{32}}{\pi^2} \cos(\phi_{v_2} - \phi_{v_3}) + \frac{8V_3 V_2 B_{32}}{\pi^2} \sin(\phi_{v_2} - \phi_{v_3}) + \frac{8V_3 V_4 G_{34}}{\pi^2} \cos(\phi_{v_4} - \phi_{v_3}) + \frac{8V_3 V_4 G_{34}}{\pi^2} \cos(\phi_{v_4} - \phi_{v_3}) \quad (6)$$

$$P_2 = \frac{8V_2^2 G_{22}}{\pi^2} + \frac{8V_2 V_1 G_{21}}{\pi^2} \cos(\phi_{v_1} - \phi_{v_2}) + \frac{8V_2 V_1 B_{21}}{\pi^2} \sin(\phi_{v_1} - \phi_{v_2}) + \frac{8V_2 V_3 G_{23}}{\pi^2} \cos(\phi_{v_3} - \phi_{v_2}) + \frac{8V_2 V_3 B_{23}}{\pi^2} \sin(\phi_{v_3} - \phi_{v_2}) + \frac{8V_2 V_4 G_{24}}{\pi^2} \cos(\phi_{v_4} - \phi_{v_2}) + \frac{8V_2 V_4 B_{24}}{\pi^2} \sin(\phi_{v_4} - \phi_{v_2}) \quad (7)$$

$$P_4 = \frac{8V_4^2 G_{44}}{\pi^2} + \frac{8V_4 V_1 G_{41}}{\pi^2} \cos(\phi_{v_1} - \phi_4) + \frac{8V_4 V_1 B_{41}}{\pi^2} \sin(\phi_{v_1} - \phi_{v_4}) + \frac{8V_4 V_2 G_{42}}{\pi^2} \cos(\phi_{v_2} - \phi_{v_4}) + \frac{8V_4 V_2 B_{42}}{\pi^2} \sin(\phi_{v_2} - \phi_{v_4}) + \frac{8V_4 V_3 G_{43}}{\pi^2} \cos(\phi_{v_3} - \phi_{v_4}) + \frac{8V_4 V_3 G_{43}}{\pi^2} \cos(\phi_{v_3} - \phi_{v_4}) \quad (8)$$

Where $B_{ij}$s are the complex component and $G_{ij}$s are the real component of the admittance matrix $Y$

The power flow between converters 1 and 3 is given by:

$$P_{13} = \frac{8V_1 V_3 G_{13}}{\pi^2} \cos(\phi_{v_3} - \phi_{v_1}) + \frac{8V_1 V_3 B_{13}}{\pi^2} \sin(\phi_{v_3} - \phi_{v_1}) \quad (9)$$

(5) and (6) are used to solve the power flow from the source converters. If the source converters' dc voltages are $V_1$ and $V_3$ and load-side single-phase converters' powers and dc voltage conversion ratio are specified as:

$P_{2min} < P_2 < P_{2max}$, $P_{4min} < P_4 < P_{4max}$,

$m_{2min} < m_{2nom} < m_{2max}$ and $m_{4min} < m_{4nom} < m_{4max}$

$$m_i = \frac{n_1 V_i}{n_i V_1} \quad i = 1,2,\dots,4$$

For a total power demand of 500W, converter 1 dc voltage of 200V and nominal dc voltage conversion ratio of 0.9, 1.1 and 1.25 for converters 2, 3 and 4 respectively, (7) and (8) and (9) are solved for the phase shift ratios of converters 2, 3 and 4. converter 1 is taken as the reference. Hence, its phase shift ratio is zero. Figure 7 shows the results of the phase shift ratios while Figure 8 shows the input powers and reactive power for total output power ranging from 0 to 500W. Despite that multiple phase shift angles of the QAB, results in figures 9 and 10 show the peak current and power at Port 1 are constant for different combinations of phase shift angles required to command same specified load power. This confirms that the single-phase-shift-controlled QAB just like the DAB, has one degree of freedom and not suitable for optimization purposes.

TABLE I. QAB PARAMETER

| Parameters | Values |
|---|---|
| Input voltage | $V_1 = 200\ V$, $V_2 = 180V$, $V_3 = 220V$, $V_4 = 250V$ |
| Phase shift ratios | $\delta_1 = 0$, $\delta_2 = 0.4560$, $\delta_3 = 0.0063$, $\delta_4 = 0.4380$ |
| Transformer winding Leakage Inductances | $L_{11} = L_{22} = L_{31} = L_{42} = 238\mu H$, $L_{12} = L_{21} = L_{32} = L_{41} = 38\mu H$ |
| Transformer winding resistances | $R_{11} = R_{22} = R_{31} = R_{42} = 0.4\Omega$, $R_{12} = R_{21} = R_{32} = R_{41} = 0.2\Omega$ |
| Turns ratio | $n_1 = n_2 = n_3 = n_4 = 2$ |
| Switching frequency | $f_s = 25kH$ |

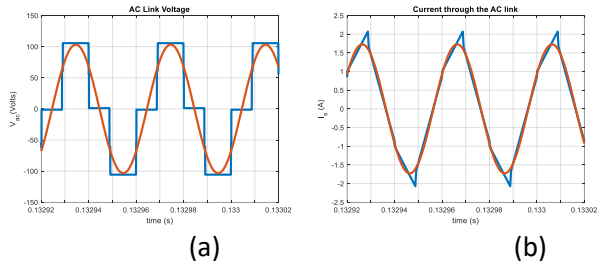

(Red - fundamental harmonic component analysis.
Blue – state-space model)
Fig. 6. Simulation results for the voltage across the ac link and current flowing through the Quad-active bridge dc-dc converter's ac link (a) Simulation results for the ac link voltage $V_{ac}$, (b) Simulation results for the ac link current

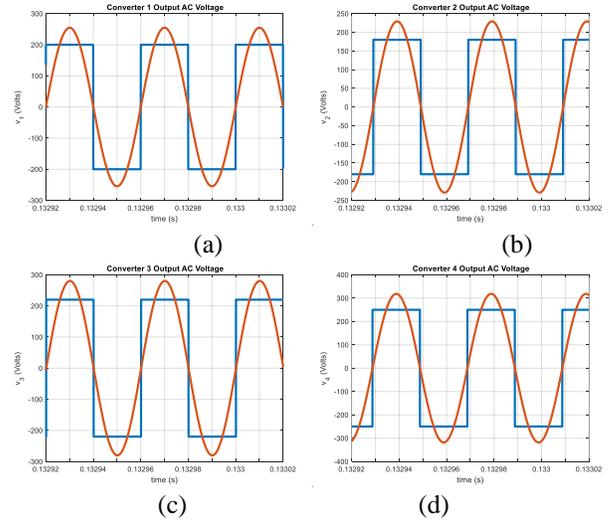

(Red - fundamental harmonic component analysis.
Blue – state-space model)
Fig. 4. Simulation results for output ac voltage of the single-phase converters in Quad-active bridge dc-dc converter. (a) converter 1 output ac voltage with a phase shift ratio of $\delta_1 = 0$; (b) converter 2 output ac $\delta_2 = 0456$; (c) Converter 3 output ac voltage $\delta_3 = 0.0063$; (d) converter 4 output ac voltage $\delta_4 = 0.4381$

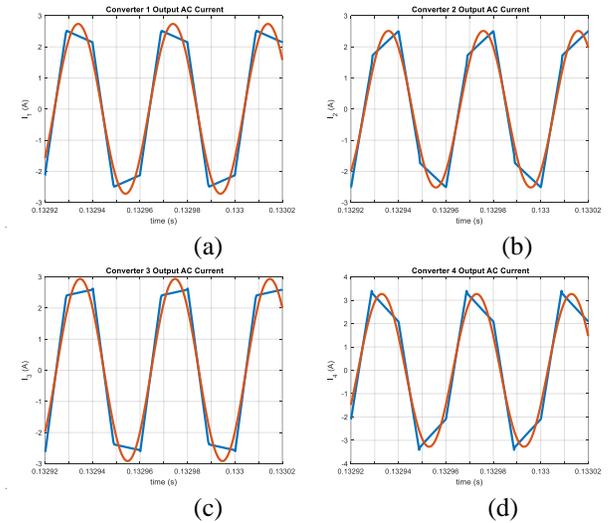

(Red - fundamental harmonic component analysis.
Blue – state-space model)
Fig. 5. Simulation results for output ac current of single-phase converters in Quad-active bridge dc-dc converter. (a) converter 1 output ac current; (b) converter 2 output ac current; (c) converter 3 output ac current; (d) converter 4 output ac current

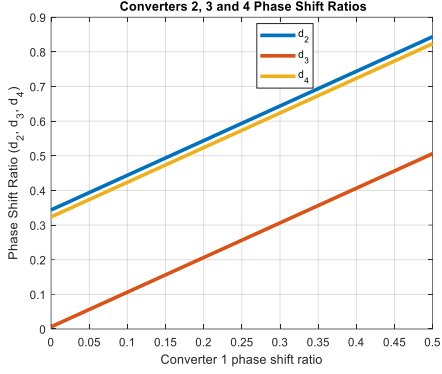

Fig. 7. Results of phase shift ratios for a load real power of 500W

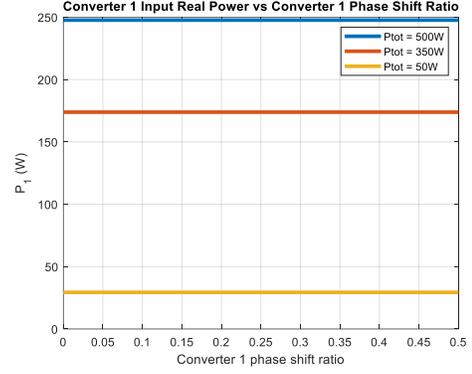

Fig. 10. converter 1 real power outputs for total load real power of 50W, 350 and 500W

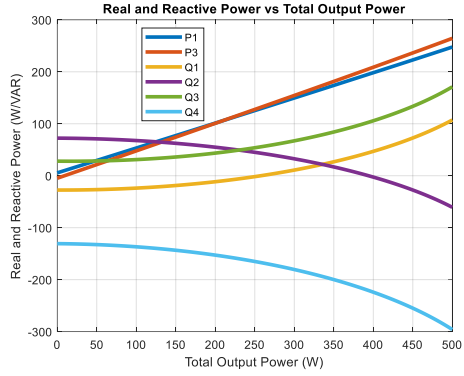

Fig. 8. Plot shows the results of converters 1 and 3 real power outputs and converters 1, 2, 3 and 4 reactive power outputs

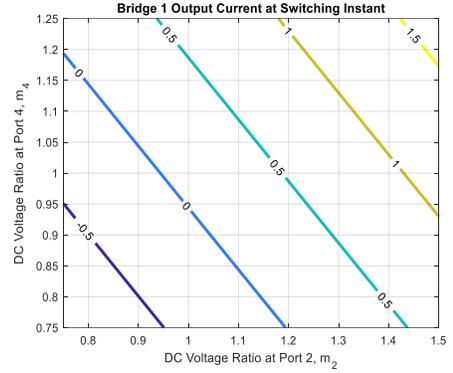

Fig. 11. The contour shows converter 1 output current at switching instant as a function of output dc voltage conversion ratio of converters 2 and 4.

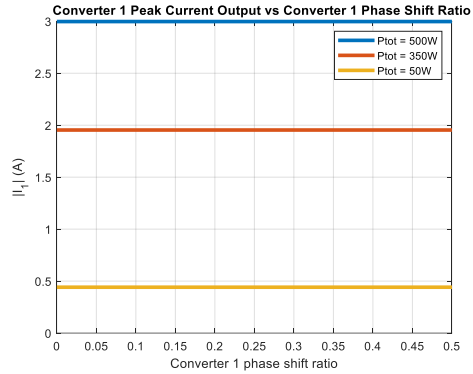

Fig. 9. Output peak current for converter 1 for total output real powers of 50W, 350W and 500W as a function of converter 1 phase shift ratio

## IV. QAB ZVS ANALYSIS

The possibility of the QAB to operate under ZVS conditions is one of its merits. It permits high frequency operation of the converter as switching losses are greatly reduced thereby improving the overall efficiency of the converter.

Phase-shift controlled full bridges will operate under ZVS conditions when the value of the full bridge output current at switching instant is negative when the direction of the current is out of the H-bridge and positive when the current flows into the bridge. The instantaneous ac currents in the four ports are given by:

$$I_1(t) = \frac{4}{\pi}\big(Y_{11}V_1 \sin(\omega_s t + \phi_{y_{11}} - \phi_{v_1}) + Y_{12}V_2 \sin(\omega_s t + \phi_{y_{12}} - \phi_{v_2}) + Y_{13}V_3 \sin(\omega_s t + \phi_{y_{13}} - \phi_{v_3}) + Y_{14}V_4 \sin(\omega_s t + \phi_{y_{14}} - \phi_{v_4})\big) \quad (10)$$

$$I_2(t) = \frac{4}{\pi}\big(Y_{21}V_1 \sin(\omega_s t + \phi_{y_{21}} - \phi_{v_1}) + Y_{22}V_2 \sin(\omega_s t + \phi_{y_{22}} - \phi_{v_2}) + Y_{23}V_3 \sin(\omega_s t + \phi_{y_{23}} - \phi_{v_3}) + Y_{24}V_4 \sin(\omega_s t + \phi_{y_{24}} - \phi_{v_4})\big) \quad (11)$$

$$I_3(t) = \frac{4}{\pi}\big(Y_{31}V_1 \sin(\omega_s t + \phi_{y_{31}} - \phi_{v_1}) + Y_{32}V_2 \sin(\omega_s t + \phi_{y_{32}} - \phi_{v_2}) + Y_{33}V_3 \sin(\omega_s t + \phi_{y_{33}} - \phi_{v_3}) + Y_{34}V_4 \sin(\omega_s t + \phi_{y_{34}} - \phi_{v_4})\big) \quad (12)$$

$$I_4(t) = \frac{4}{\pi}\big(Y_{41}V_1 \sin(\omega_s t + \phi_{y_{41}} - \phi_{v_1}) + Y_{42}V_2 \sin(\omega_s t + \phi_{y_{42}} - \phi_{v_2}) + Y_{43}V_3 \sin(\omega_s t + \phi_{y_{43}} - \phi_{v_3}) + Y_{44}V_4 \sin(\omega_s t + \phi_{y_{44}} - \phi_{v_4})\big) \quad (13)$$

For the QAB, the ZVS boundary conditions are:

Port 1: $I_1(\delta_1 T_h) < 0$. Port 2: $I_2(\delta_2 T_h) > 0$.

Port 3: $I_3(\delta_3 T_h) < 0$. Port 4: $I_4(\delta_4 T_h) > 0$.

$T_h$ is half of the switching period. The voltages of the source converters are fixed and the dc voltage ratios for the load side converters vary between 0.8 to 1.5 for converter 2 and 0.75 to 1.25 for converter 4. converter 1 at 10% of total rated output power, meets the ZVS conditions in the region to the left the zero line in Fig. 11. This shows the range of values of the load-side converter dc voltages that meets the ZVS requirements at the specified load power. The result in Figure 12 confirms this observation as it is the same region that produces positive reactive power required for ZVS at converter 1. This means input ac current lags the voltage which is another definition for ZVS conditions in single phase converters.

A similar explanation of the result obtained for converter 1 holds for converter 3. For converter 2, the region which meets the ZVS requirement is to the right of the zero line in Fig 13. This is the region when the current at the switching instant is positive, and the output reactive power is negative as shown in Figure 14. Here, the ac current leads the voltage. When the input dc voltage of the source converters 1 and 3 and the dc voltage of converter 2 are kept constant, and the dc voltage conversion ratio of Port 4 varies from 0.75 to 1.25, while the total demanded load power ranges from 0 to 380W, the entire region of the plot in Figure 15 meets the ZVS region as the current at the switching instant is negative in this region. As confirmed in Figure 16, the reactive power is also positive throughout this region. This indicate the ac current is lagging the voltage. This result is like the one obtained for converter 1. The switches in converter 4 will meet the ZVS requirement in the region above the zero curve, in Figure 17. In this region the switching current is negative, and the reactive power is positive (see Figure 18) which indicate the ac current is leading the Voltage.

## V. EXPERIMENTAL VERIFICATION

To verify the foregoing analysis, an experiment was conducted. For the experiment, dc voltages of 30.86, 26.07, 30.72, and 27.3V, were applied to converters 1,2,3 and 4 respectively. At high output power, there is usually enough circulating current to fully charge the output capacitance of the switches within the dead-time, thereby causing the switches to turn on at zero voltage. Hence, the issue with ZVS common at low or close to zero output power. For load power as low as 15W, the instantaneous switching currents are -0.5336A, -0.5754A, -0.5110A and -0.3416A for converters 1, 2, 3, and 4 respectively According to the ZVS boundary conditions, converters 1 and 3 meets the ZVS requirements while converters 2 and 4 do not. Again, a simulation of the quad active bridge dc-dc converter was carried out for the stated dc voltages and phase shift ratios of 0, 0.0314, 0.0053 and 0.030 for the single-phase converters 1, 2, 3 and 4 respectively, of the quad active bridge converters. The simulation results for converters 1 and 3 confirm the result from the ZVS analysis. It is shown from the plot in Figures 19 and 21 that the output current of converter 1 and 3 lags its output voltage. The experimental result shown in Figure 23 further confirms the result from the ZVS analysis. Also, from the results of the ZVS analysis for converters 2 and 4, the current at switching instants are negative. For these converters to meet the ZVS requirement, the currents should be positive. Hence, the converters do not meet the ZVS conditions. This result is further confirmed by the simulation and experimental results shown in Figures 20, 22 and 24.

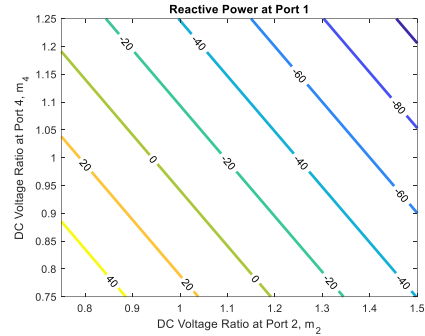

Fig. 12. The contour shows reactive power output of converter 1 as a function of output dc voltage conversion ratio of converters 2 and 4.

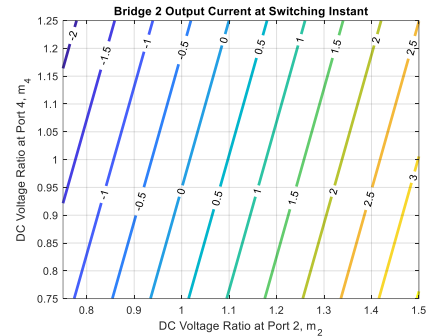

Fig. 13. The contour shows converter 2 output current at switching instant as a function of output dc voltage conversion ratio of converters 2 and 4.

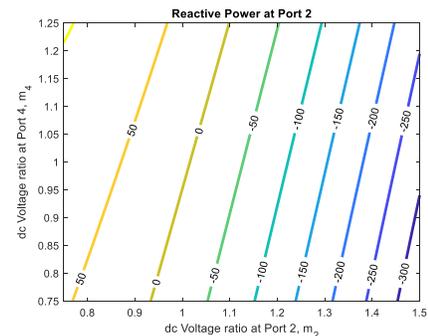

Fig. 14. The contour shows reactive power output of converter 2 as a function of output dc voltage conversion ratio of converters 2 and 4.

Fig. 15. The contour shows converter 3 output current at switching instant as a function of output dc voltage conversion ratio of converter 4 and total load power.

Fig. 16. The contour shows the reactive power output of converter 3 as a function of output dc voltage conversion ratio of converter 4 and total load power.

Fig. 17. The contour shows converter 4 output current at switching instant as a function of output dc voltage conversion ratio of converter 4 and total load power.

Fig. 18. The contour shows the reactive power output of converter 4 as a function of output dc voltage conversion ratio of converter 4 and total load power.

Fig. 19. Simulation result for converter 1 output current and voltage at dc voltages of 30.86V, 26V, 30.72V and 27V for Converters 1, 2, 3 and 4 respectively and 15W total power

Fig. 20. Simulation result for converter 2 output current and voltage at dc voltages of 30.86V, 26V, 30.72V and 27V for converters 1, 2, 3 and 4 respectively and 15W total power

Fig. 21. Simulation result for converter 3 output current and voltage at dc voltages of 30.86V, 26V, 30.72V and 27V for converters 1, 2, 3 and 4 respectively and 15W total power

Fig. 22. Simulation result for converter 4 output current and voltage at dc voltages of 30.86V, 26V, 30.72V and 27V for converters 1, 2, 3 and 4 respectively and 15W total power

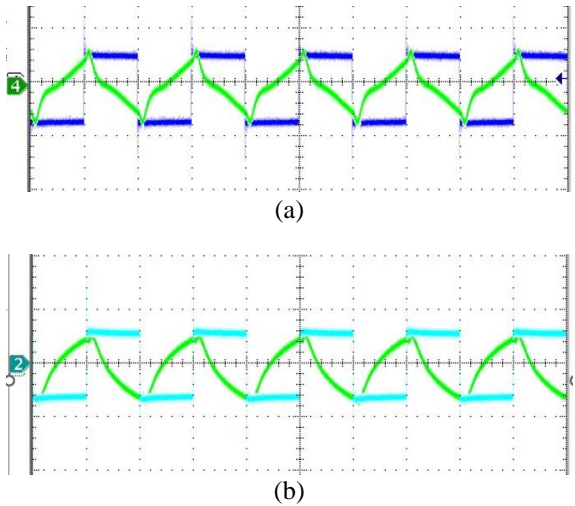

Fig. 23. Experimental result for converters 1 and 3 output current and voltage at dc voltages of 30.86V, 26V, 30.72V and 27V for converters 1, 2, 3 and 4 respectively and 15W total power output (a) converter 1 output voltage at 50V/div in blue and output current at 2A/div in green. (b) converter 3 output voltage at 50V/div in cyan and output current at 5A/div in green.

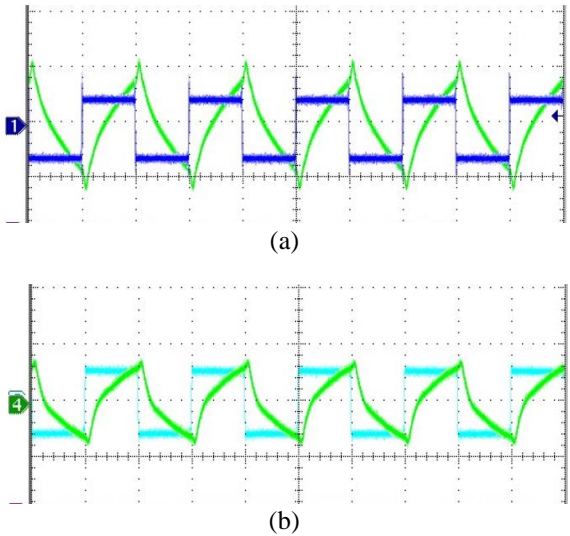

Fig. 24. Experimental result for converters 2 and 4 output current and voltage at dc voltages of 30.86V, 26V, 30.72V and 27V for converters 1, 2, 3 and 4 respectively and 15W total power output (a) converter 2 output voltage at 50V/div in blue and output current at 2A/div in green. (b) converter 4 output voltage at 50V/div in cyan and output current at 5A/div in green.

## VI. CONCLUSION

The power flow analysis of a Quad-active bridge dc-dc converter was carried out using the full equivalent circuit model as opposed to the lossless start-equivalent model of the converter. From the power flow analysis of the converter, it was observed that one solution point exists for different combinations of phase shift angles for a SPS-modulated QAB. Hence optimization of the control variables may not be possible for this modulation technique for a MAB converter. From the ZVS analysis, the range of values of the dc voltage ratio that ensures ZVS at specified output power for the different full bridges were also determined. From the ZVS analysis the load voltage (that lie between possible range of voltages) that ensures ZVS may be determined.

REFERENCES

[1] L. Ortega, P. Zumel, C. Fernandez, J. Lopez-Lopez, A. Lazaro, and A. Barrado, "Power Distribution Algorithm and Steady-State Operation Analysis of a Modular Multiactive Bridge Converter," *IEEE Trans. Transp. Electrif.*, vol. 6, no. 3, pp. 1035–1050, 2020, doi: 10.1109/TTE.2020.3010003.

[2] A. Chandwani and A. Mallik, "Three-loop Multi-variable Control of Triple Active Bridge Converter with Power Flow Optimization," in *Conference Proceedings - IEEE Applied Power Electronics Conference and Exposition - APEC*, 2022, pp. 2008–2013, doi: 10.1109/APEC43599.2022.9773652.

[3] M. Neubert, A. Gorodnichev, J. Gottschlich, and R. W. De Doncker, "Performance analysis of a triple-active bridge converter for interconnection of future dc-grids," *ECCE 2016 - IEEE Energy Convers. Congr. Expo. Proc.*, 2016, doi: 10.1109/ECCE.2016.7855337.

[4] P. Purgat, S. Bandyopadhyay, Z. Qin, and P. Bauer, "Zero Voltage Switching Criteria of Triple Active Bridge Converter," *IEEE Trans. Power Electron.*, vol. 36, no. 5, pp. 5425–5439, 2021, doi: 10.1109/TPEL.2020.3027785.

[5] S. Bandyopadhyay, Z. Qin, and P. Bauer, "Decoupling Control of Multiactive Bridge Converters Using Linear Active Disturbance Rejection," *IEEE Trans. Ind. Electron.*, vol. 68, no. 11, pp. 10688–10698, 2021, doi: 10.1109/TIE.2020.3031531.

[6] N. Naseem and H. Cha, "Quad-Active-Bridge Converter with Current Balancing Coupled Inductor for SST Application," *IEEE Trans. Power Electron.*, vol. 36, no. 11, pp. 12528–12539, 2021, doi: 10.1109/TPEL.2021.3076460.

[7] S. Falcones, R. Ayyanar, and X. Mao, "A DC-DC Multiport-converter-based solid-state transformer integrating distributed generation and storage," *IEEE Trans. Power Electron.*, vol. 28, no. 5, pp. 2192–2203, 2013, doi: 10.1109/TPEL.2012.2215965.